\documentclass[12pt,preprint]{aastex}

\shorttitle{}
\shortauthors{Endl et al.}

\begin{document}

\title{Revisiting $\rho^{1}$~Cancri~e: A New Mass Determination Of The Transiting super-Earth  
\footnote{Based partly on observations obtained with the Hobby-Eberly Telescope, which is a joint project of the 
University of Texas at Austin, the Pennsylvania State University, Stanford University, 
Ludwig-Maximilians-Universit\"at M\"unchen, and Georg-August-Universit\"at G\"ottingen.}}

\author{Michael Endl}
\affil{McDonald Observatory, The University of Texas at Austin,  
    Austin, TX 78712}
\email{mike@astro.as.utexas.edu}
\author{Paul Robertson}                       
\affil{Department of Astronomy, The University of Texas at Austin,
    Austin, TX 78712}
\author{William D. Cochran}                       
\affil{McDonald Observatory, The University of Texas at Austin,
    Austin, TX 78712}
\author{Phillip J. MacQueen}                       
\affil{McDonald Observatory, The University of Texas at Austin,
    Austin, TX 78712}
\author{Erik J. Brugamyer}
\affil{Department of Astronomy, The University of Texas at Austin,
    Austin, TX 78712}
\author{Caroline Caldwell}
\affil{Department of Astronomy, The University of Texas at Austin,
    Austin, TX 78712}
\author{Robert A. Wittenmyer}
\affil{Department of Astrophysics and Optics, School of Physics, University of New South Wales, Sydney, Australia}
\author{Stuart I. Barnes}
\affil{McDonald Observatory, The University of Texas at Austin,
    Austin, TX 78712}
\author{Kevin Gullikson}
\affil{Department of Astronomy, The University of Texas at Austin,
    Austin, TX 78712}

\begin{abstract}

We present a mass determination for the transiting super-Earth $\rho^{1}$~Cancri~e
based on nearly 700 precise radial velocity (RV) measurements. This extensive
RV data set consists of data collected by the McDonald Observatory planet search 
and published data from Lick and Keck observatories (Fischer et al.~2008). 
We obtained 212 RV measurements with the Tull Coud\'e Spectrograph at the Harlan J. Smith 2.7\,m Telescope and  
combined them with a new Doppler reduction of the 131 spectra that we 
have taken in 2003-2004 with the High-Resolution-Spectrograph (HRS) at the Hobby-Eberly Telescope (HET) for the
original discovery of $\rho^{1}$~Cancri~e. 
Using this large data set we obtain a 5-planet Keplerian orbital solution for the system and 
measure an RV semi-amplitude of $K=6.29\pm0.21$\,m\,s$^{-1}$ for $\rho^{1}$~Cnc~e and  
determine a mass of 8.37$\pm0.38$~M$_{\oplus}$. The uncertainty in mass is thus less than $5\%$.
This planet was previously found to transit its parent star (Winn et al.~2011,
Demory et al.~2011), which allowed them to estimate its radius. Combined with the latest
radius estimate from Gillon et al.~(2012), we obtain a mean density of 
$\rho=4.50\pm0.20$\,g\,cm$^{-3}$. The location of  
$\rho^{1}$~Cnc~e in the mass-radius diagram suggests that the planet contains a significant amount of volitales, 
possibly a water-rich envelope surrounding a rocky core.

\end{abstract}


\section{Introduction}

The $\rho^{1}$~Cancri planetary system is one of the most interesting, nearby multi-planet systems that was discovered and extensively studied by the
radial velocity (RV) technique. The parent star, $\rho^{1}$~Cancri (= HR~3522, HD~75732, 55~Cancri), is a $V=5.95$ G8V (Montes et al.~2001) star, located at a distance of $12.3\pm0.1$\,pc, 
based on the Hipparcos parallax of $81.03\pm0.75$~mas (Van Leeuwen~2007). It is the primary of a wide visual binary with a projected separation of $\approx1065$~AU (Mugrauer et al. 2006).
The first planet in this system ($\rho^{1}$~Cnc~b, $P=14.65$~d) was found by Butler et al.~(1996), based on Lick Observatory RV data. Six years later, 
Marcy et al.~(2002) presented evidence for two more giant planets orbiting this star: $\rho^{1}$~Cnc~c with a period of 44~d and a very long period planet at an orbital separation 
of 5.5~AU, $\rho^{1}$~Cnc~d. An intensive RV campaign that we carried out using the Hobby-Eberly Telescope (HET), revealed a short-periodic signal that 
was, at that time, thought to be one of the first discoveries of a hot Neptune with a minimum mass of $\sim 17$~M$_{\oplus}$ and an orbital period of 2.8 days (McArthur et al. 2004, 
hereafter Mc04), raising the total number of detected planets in this system to four. The planet count was further increased by Fischer et al.~(2008, hereafter F08), who 
presented evidence for a fifth planet with an orbital period of 260~days. 

Dawson \& Fabrycky (2010, hereafter DF10) re-analyzed the published RV time-series data for this star and claimed that the 2.8~day period of $\rho^{1}$~Cnc~e is an alias and that 
the true period of this companion is just 0.7365 days. This shorter period also led to a reduction of the minimum mass of $\rho^{1}$~Cnc~e to around $8$~M$_{\oplus}$, moving 
the planet from the Neptune mass range into the super-Earth mass bin. DF10 pointed out the high a-priori transit probability of 25\% for a planet 
and motivated highly precise photometric observations to search for the planetary transit signal. This transit search was indeed successful, using two different space 
telescopes, MOST and warm {\it Spitzer}. Winn et al.~(2011, hereafter W11), and Demory et al.~(2011, hereafter D11) practically simultaneously announced the detection of the transit 
signal of this super-Earth. The transit thus confirmed the shorter orbital period for $\rho^{1}$~Cnc~e, as 
suggested by DF10. D11 measure a planetary radius of $2.1\pm0.17~R_{\oplus}$ and 
a mean density of $4.8\pm1.3$\,g\,cm$^{-3}$, while W11 reported a similar radius of $2.0\pm0.14~R_{\oplus}$ and slightly higher mean density of $5.9^{+1.5}_{-1.1}$\,g\,cm$^{-3}$. 
This places $\rho^{1}$~Cnc~e in the mass-radius diagram between the region of high-density, rocky planets like CoRoT-7b (Hatzes et al. 
2011) and Kepler-10b (Batalha et al.~2011) and planets with a significant amount of volatiles, so-called ``Mini-Neptunes'', like GJ~1214b (Charbonneau et al.~2009)
and the Kepler-11 planets (Lissauer et al.~2011).   
An improved radius determination of $2.17\pm0.10~R_{\oplus}$ was presented by Gillon et al.~(2012, hereafter G12) by combining the {\it Spitzer} with the MOST photometry. Even more
recently, Demory et al.~(2012) used Warm {\it Spitzer} 4.5~$\mu$m observations of occultations of $\rho^{1}$~Cnc~e to detect its thermal emission. Clearly, this nearby transiting
super-Earth planet around a bright star offers an abundant variety of very interesting follow-up observations that will allow a detailed characterization of this exoplanet.   

In our paper we will focus on the mass determination for $\rho^{1}$~Cnc~e based on hundreds of precise RV measurements.   
The paper is structured as follows: we first describe the observations of $\rho^{1}$~Cnc at McDonald Observatory, the second section contains a description of the multi-planet 
orbital fit that we have performed, and the third section discusses our results, in particular the precise mass, for $\rho^{1}$~Cnc~e. 

\section{McDonald Observatory Observations} 

We observed $\rho^{1}$~Cancri as part of the long-term Doppler exoplanet survey at the Harlan J. Smith 2.7\,m Telescope 
(HJST) (e.g. Cochran et al.~1997, Robertson et al.~2012) beginning 
in May 1999. For all observations we used the Tull Coud\'e Spectrograph (Tull et al.~1995) in combination with a $1.2^{''}$ slit that yields
a resolving power of $R=60,000$ with 2-pixel sampling. The star light passes through a 
temperature controlled iodine vapor (I$_2$) cell that is mounted in front of the entrance slit.
The I$_2$-cell provides a precise wavelength calibration and allows the reconstruction of the instrumental
profile at the time of observation. Precise differential RVs are computed from each spectrum using our {\it Austral} Doppler code (Endl et al.~2000).

As a consequence of the claim of DF10 that the true period of the inner planet is just 0.7365~days, we changed our observing
strategy for this star to test this prediction. While $\rho^{1}$~Cancri was previously observed with a cadence of once a month or less, we
increased the cadence dramatically in 2010. To allow a good sampling of such a short period we typically observed $\rho^{1}$~Cancri several times per
month and often up to three times per night.    
Through June 2012 we collected a total of 212 single exposures, expanding
the total time baseline of the HJST RV data to 4728 days or almost 13 years.

We combine these RV measurements with an improved Doppler-reduction of the 131 spectra obtained with the High-Resolution-Spectrograph (HRS) (Tull et al.~1995) at the
Hobby-Eberly Telescope (HET) (Ramsey et al. 1998) over the time span of 190 days from October 2003 to April 2004.  

The 343 RV results from HJST/Tull and HET/HRS are displayed in Figure\,\ref{rvs}  
and listed in Table\,\ref{rvs}. The large scatter in the RV results shown in Figure\,\ref{rvs} is almost entirely due to Keplerian motion caused by the
planetary system.    

\section{Keplerian Orbital Solutions \& Mass Determination}
\label{gf}

We used GaussFit (Jefferys et al. 1988) to perform a simultaneous fit to our RV data of the known four
giant planets in this system with periods of 14.7, 43, 260, and $\approx4700$ days. We do not add an additional noise term, often
called ``jitter'', to the uncertainties of the RV measurements. Jitter is sometimes used to account for instrinsic stellar variability, 
algorithmical noise and/or residual instrumental systematic effects (Wright 2005). However, as the physical cause is poorly understood
we decided not to add any jitter to the RV data. In principle, also the presence of additional planets can be a cause of
larger residual scatter.   

The 4-planet solution yields periods of P1=$14.6514\pm0.0002$~d, P2=$43.0\pm0.01$~d, P3=$261.8\pm0.3$~d and P4=$4660\pm37$~d. 
The total rms-scatter of the 343 RV measurements around the 4-planet model fit is 7.5\,m\,s$^{-1}$ (HJST/Tull: 7.7\,m\,s$^{-1}$, 
HET/HRS: 7.2\,m\,s$^{-1}$). The reduced $\chi^{2}$ of this fit is 7.6.
The biggest difference
to F08 is the shorter period for the outermost gas giant. However, the F08 period of $5218 \pm 230$~d is longer than the
time baseline of our observations which probably explains why we do not find this longer period.
We tested this by including the Lick data in the 4-planet model, which yield $4929 \pm 45$~d for the best-fit period of planet d.

\begin{figure}
\includegraphics[angle=270,scale=.62]{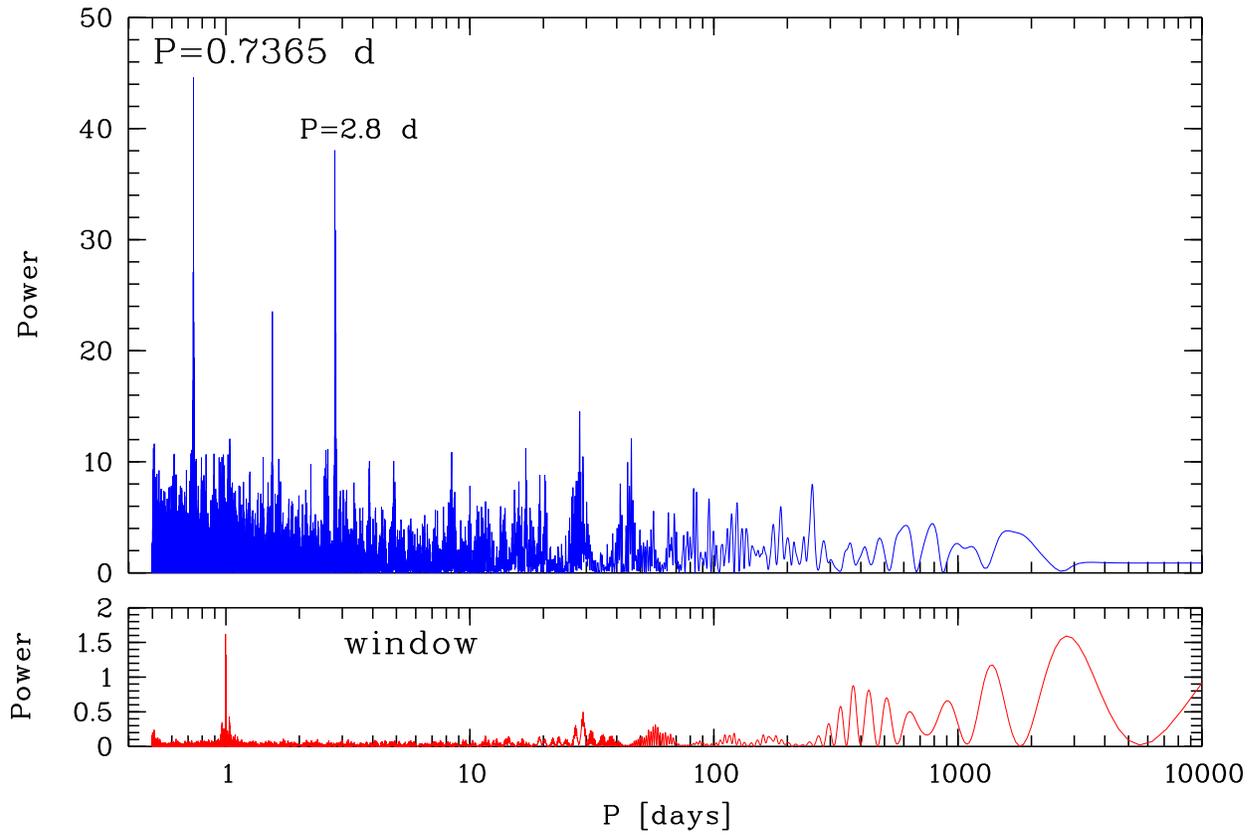}
\caption{Lomb-Scargle periodogram of the 343 RV residuals after subtracting the orbits of the 4 known giant planets.
Clearly, the shorter period at 0.74~d, as suggested by DF10, is a stronger peak than the alias at 2.8~d. 
\label{per}}
\end{figure}

We then searched the McDonald Observatory RV residuals from the 4-planet fit for a periodic signal due to the inner-most companion. 
The Lomb-Scargle periodogram of the residuals is displayed in Fig.\,\ref{per} and has two prominent and highly significant peaks. 
The less significant peak with a Scargle-power of 38 is at a period of $2.8$~days, the original period of planet e announced by M04.
The shorter period of 0.7365~days is clearly the stronger peak with a power of 44.6.

For all subsequent orbital fits we fixed the eccentricity of $\rho^{1}$~Cnc~e to $0$ because nearly circular orbits are expected
based on tidal damping (DF10) and from classic secular theory (Van Laerhoven \& Greenberg 2012). The observed 
occulation phase from D12 also sets an upper limit of $e<0.06$ at the $3\sigma$ level, consistent with
this expectation.

In order to derive a precise companion mass from our measured $K$ amplitude we also need a precise mass for the host star. 
We adopt the latest value of $0.905\pm0.015~M_{\odot}$ from von Braun et al.~(2011).

We then fit a 5-planet model to our RV data with $P=0.7365$~d for the inner-most companion. 
The 5-planet Keplerian fit yields a reduced $\chi^{2}$ of 3.8 and the residual rms-scatter around the
fit is now reduced to 5.3\,m\,s$^{-1}$ (HJST/Tull: 5.5\,m\,s$^{-1}$, 
HET/HRS: 5.2\,m\,s$^{-1}$). For $\rho^{1}$~Cnc~e we obtain an RV semi-amplitude of 
$K=6.01\pm0.38$\,m\,s$^{-1}$. The errors produced by GaussFit are generated from a maximum likelihood 
estimation that is an approximation to a Bayesian maximum a posteriori estimator with a flat prior (Jefferys~1990). 
Using the McDonald data alone derive a minimum mass of $7.95\pm0.57~M_{\oplus}$ for $\rho^{1}$~Cnc~e. 
The stellar mass uncertainty of 1.7\% and the $K$ amplitude error of 6.2\% hence lead to a planetary mass (1-$\sigma$) uncertainty of 7.2\%. 
  
To further improve the precision of this mass determination we add the 70 published RV measurements from Keck Observatory from F08. 
The Keck data have a time baseline of 2037 days.
We performed a new 5-planet fit including the abritrary RV zero-point of the Keck data set as a free parameter. The 
residual
scatter of the 413 RV points is 5.5\,m\,s$^{-1}$ (HJST/Tull: 6.1\,m\,s$^{-1}$, 
HET/HRS: 5.6\,m\,s$^{-1}$, and Keck/HIRES: 3.1\,m\,s$^{-1}$). Again, we do not detect any significant signal in the
RV residuals. 
The HJST+HET+Keck 5-planet fit yields $K=6.22\pm0.24$\,m\,s$^{-1}$ and 
a minimum mass of $8.20\pm0.41~M_{\oplus}$. 
The addition of the Keck data increases the precision of the planet's mass determination to 5\%.   

\begin{figure}
\includegraphics[angle=270,scale=.62]{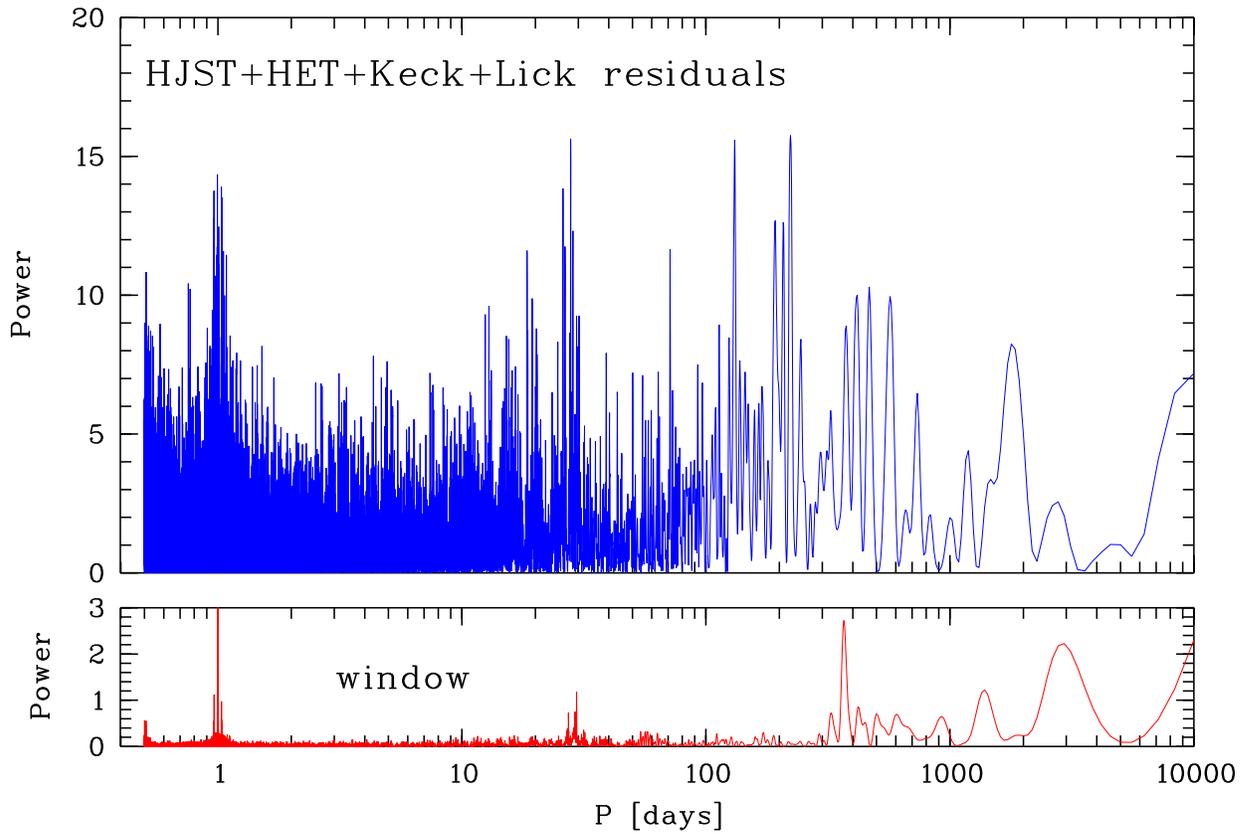}
\caption{Lomb-Scargle periodogram of the 663 RV residuals after subtracting the orbits of the 5 planets in the
$\rho^{1}$~Cnc system. There is no significant signal apparent that could indicate the presence of a 6th planet
in the system.
\label{resper}}
\end{figure}

In a final step we also include the published long-term (6642 days) RV data (250 measurements) from Lick Observatory from F08. 
This raises the total number of precise RV measurements to 663 and the total time coverage to 8476 days or 23.2 years. The 
final 5-planet model using HJST+HET+Keck+Lick has a residual rms-scatter of 6.3\,m\,s$^{-1}$ (HJST/Tull: 6.1\,m\,s$^{-1}$, 
HET/HRS: 5.6\,m\,s$^{-1}$, Keck/HIRES: 3.3\,m\,s$^{-1}$ and Lick/Hamilton: 7.6\,m\,s$^{-1}$) . 
The Lomb-Scargle periodogram of
the residuals shows no significant peak rising over the noise (see Figure\,\ref{resper}). Despite over 600 precise RV measurements we do not have
any indication yet for a 6th planet in the system. 

\begin{figure}
\includegraphics[angle=270,scale=.60]{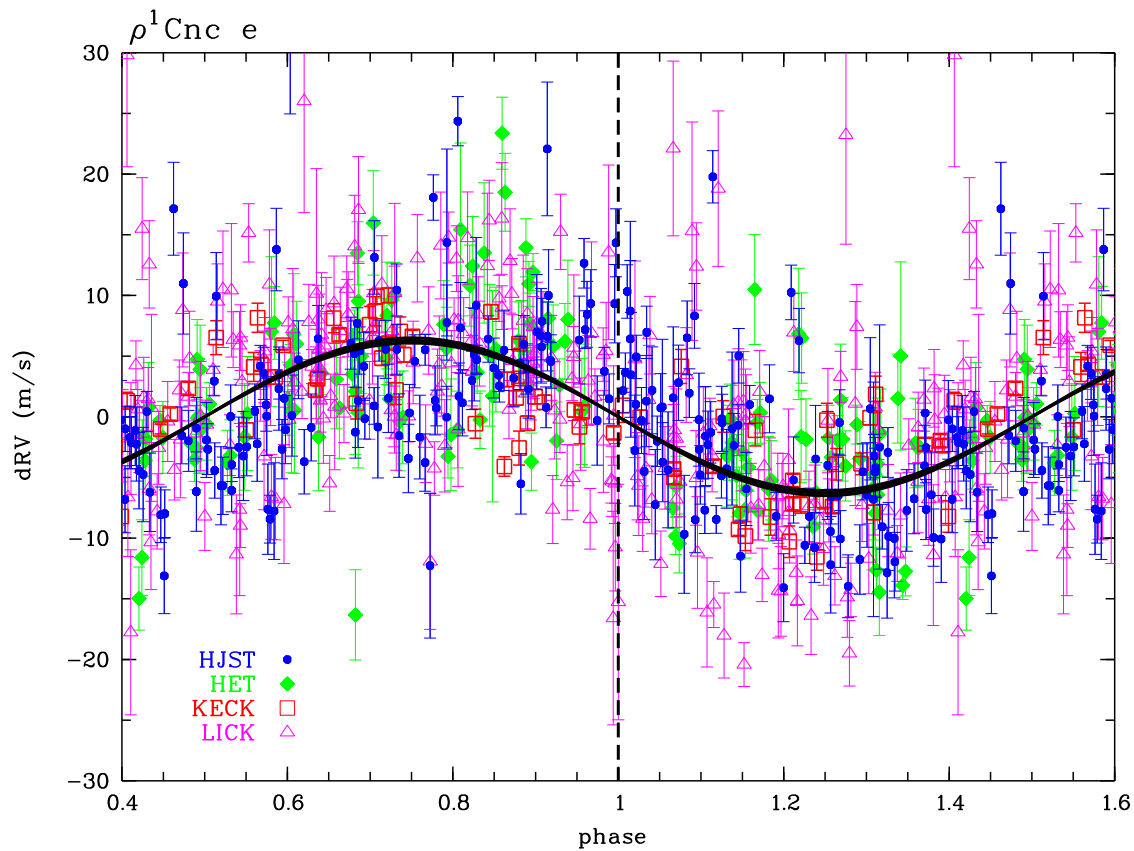}
\caption{The 663 RV measurements of $\rho^{1}$~Cnc from HJST (filled circles), HET (diamonds), Keck (boxes) and 
Lick (triangles), after subtracting the four RV-orbits of the giant planets and
phased to the transit ephemeris of D11 (transit occurs at phase 1, indicated by the vertical dashed line).
Our best-fit Keplerian orbit for the super-Earth is shown as solid line. The width of this line represents
the 1-$\sigma$ uncertainty of 0.21\,m\,s$^{-1}$ in the RV semi-amplitude $K$.
The residual scatter around this orbit is 6.3\,m\,s$^{-1}$ (some of the worst outliers are outside the RV range shown here).
 \label{orbit}}
\end{figure}

Including the Lick data changes the value for the semi-amplitude of $\rho^{1}$~Cnc~e slightly to
$K=6.29\pm0.21$\,m\,s$^{-1}$ and the minimum mass to $8.30\pm0.38~M_{\oplus}$. 
By using the McDonald data together with the Keck and Lick RVs we thus
arrive at a mass uncertainty of 4.6\%.   
The phase-folded RV data are displayed in Fig.\ref{orbit} along with the best-fit Keplerian solution for $\rho^{1}$~Cnc~e after
subtracting the four other planets. 
Table\,\ref{super} summarizes the parameters we obtained for $\rho^{1}$~Cnc~e from this full 5-planet solution. The orbital
parameters for the four giant planets in the $\rho^{1}$~Cnc system are listed in Table\,\ref{4planets}. Since we use the 
entire RV data, that were accumulated by the McDonald Observatory and California planet search, these orbital parameters should
also be the best values to the 5-planet system. 

\section{Discussion}

With the inclination $i$ of 82.5 degrees measured by G12 from the photometric transit, we derive a planetary mass of
$8.37\pm0.38~M_{\oplus}$ for $\rho^{1}$~Cnc~e. This value is slightly higher than reported by D11 ($7.81\pm0.56~M_{\oplus}$) and slightly lower 
than the value of W11 ($8.63\pm0.35~M_{\oplus}$). The W11 mass and its error were taken from the orbital solutions of DF10. 
With our approach, we could not reproduce the 
small error in $K$ of only 0.2\,m\,s$^{-1}$ presented by DF10 using the Keck and Lick RV data alone (we obtain an
uncertainty in $K$ of 0.33\,m\,s$^{-1}$). 

Our mass estimate is based on
twice the amount of RV data and we also benefit from the improved stellar 
mass determination of von Braun et al.~(2011). 
We therefore regard this new mass determination as the current best 
value for $\rho^{1}$~Cnc~e. The uncertainty in mass is driven by the uncertainty in $K$ and in second order coupled to the uncertainty 
in the mass of the host star.
We are thus only limited by the precision of the radial velocity measurements and the quality of the stellar parameters of $\rho^{1}$~Cnc. 
 
\begin{figure}
\includegraphics[angle=0,scale=.60]{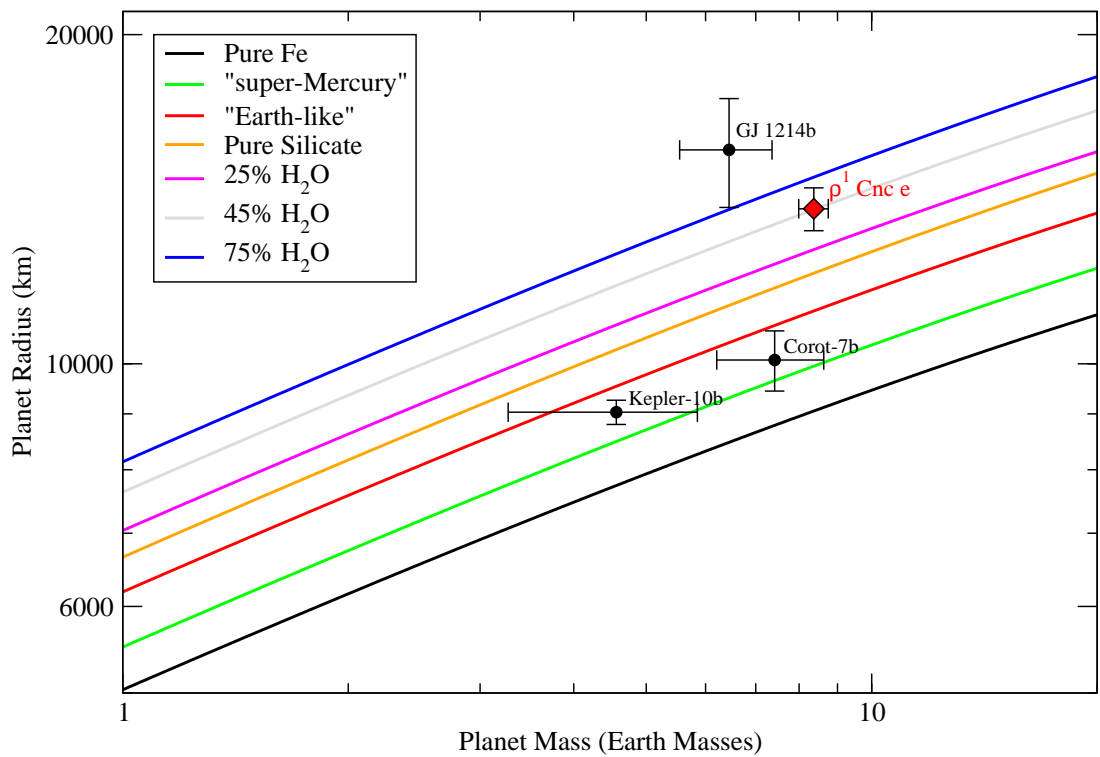}
\caption{
$\rho^{1}$~Cnc e in the mass-radius diagram and compared to other transiting super-Earths.
The radius value was taken from G12 and the mass is from this work. The parameters for
CoRoT-7b were adopted from Hatzes et al.~(2011) and Kepler-10b from Batalha et al.(2011). 
The radius value of GJ~1214b represent an average of the two estimates presented in Carter et al.~(2011).
The models for the internal composition of planets in this mass and radius range are taken from Seager et al.~(2007).
\label{mr}}
\end{figure}

Figure\,\ref{mr} shows $\rho^{1}$~Cnc~e in the mass-radius diagram compared to models for internal composition of 
small and low-mass planets from
Seager et al.~(2007) and three other transiting planets with well determined masses and radii. $\rho^{1}$~Cnc~e 
has the smallest area of uncertainty based on its errors in mass and radius (Kepler-10b has a smaller 
radius error but a larger uncertainty in mass). 
As noted by other authors (e.g. G11), $\rho^{1}$~Cnc~e
requires a significant amount of volatiles to explain its location in this diagram. It is located
significantly above the model curves for purely rocky planets and approaches the zone of ``mini-Neptunes''.    
Still, it does not require a large H/He envelope, but its mass and radius rather suggest a water-rich envelope around a 
rocky core. 

Kaib, Raymond \& Duncan~(2011) suggest that the $\rho^{1}$~Cnc planetary system is coplanar but misaligned with its host star spin axis due to
the perturbations of the secondary star. In principle, this can be tested (at least for $\rho^{1}$~Cnc~e) by observing and
measuring the Rossiter-McLaughlin (RM) effect. 
However, the expected 
amplitude of the RM effect for $\rho^{1}$~Cnc~e is only 0.5\,m\,s$^{-1}$, and while RV measurements 
were obtained during transit by coincidence, a signal of this small amplitude is
clearly undetectable by the current data. Future extreme precision 
RV measurements of several transits (by e.g. the upgraded HET/HRS or 
HARPS-North) might allow us to measure the spin-orbit
misalignment for this planet.
    
As mentioned before, despite the large quantity of precise RV measurements
we did not detect any significant residual signal that could indicate a sixth planet in the system.
The most interesting peak is near 131 days, as the window function is clean at this
period value, and it would be at an orbital separation between the 43~d and the 261~d planet.
But this peak has only a modest power and is not statistically signifcant. 
However, there is a rapidly increasing ``treasure trove'' of precise RV measurements for this system, 
with our paper adding over 300 RV points to
the published sample. This should allow in the future to achieve sensitivity for more planets, either with lower mass or at longer orbital 
periods and especially in the habitable zone and in the currently large empty region
between the inner
4 planets and the distant outer planet at $a\approx5$~AU. The $\rho^{1}$~Cnc multi-planetary system likely has more exciting discoveries 
waiting to be made.
 
\acknowledgments
We thank our referee Kaspar von Braun for his thoughtful comments that helped
to improve the manuscript.
This material is based on work supported by the National Aeronautics and
Space Administration under Grant NNX09AB30G through the Origins of Solar Systems program.
P. Robertson is supported by a University of Texas at Austin Continuing Fellowship.
The Hobby-Eberly Telescope (HET) is a joint project of the University of
Texas at Austin, the Pennsylvania State University, Stanford University,
Ludwig-Maximilians-Universit\"{a}t M\"{u}nchen,
and Georg-August-Universit\"{a}t G\"{o}ttingen.
The HET is named in honor of its principal benefactors,
William P. Hobby and Robert E. Eberly. We would like to thank the
McDonald Observatory TAC for generous allocation of observing time.
We are grateful to the HET Resident Astronomers and Telescope Operators for their valueable 
assistance in gathering our HET/HRS data. We also thank all previous observers of
the planet search program at the HJST: Artie P. Hatzes, Diane Paulson and Candece Gray.
We also would like to thank the California planet search group for publishing their extensive 
RV data from Lick and Keck observatories.


\begin{deluxetable}{lrrr}
\tablecolumns{4}
\tablewidth{0pt}
\tablecaption{Precise Radial Velocity Results
\label{rvs}}
\tablehead{
\colhead{BJD} & {dRV(m\,s$^{-1}$)} & {err(m\,s$^{-1}$)} & {Telescope/Spectrograph}
}
\startdata
51326.629453 & -22567.53 & 11.8 & HJST/Tull\\
51503.889809 & -22568.26 &  2.7 & HJST/Tull\\
51529.976108 & -22626.99 &  1.3 & HJST/Tull\\ 
.... & & \\
\hline
\enddata
\end{deluxetable}

\begin{deluxetable}{lrr}
\tablecolumns{5}
\tablewidth{0pt}
\tablecaption{Keplerian Orbital Solution for the super-Earth planet $\rho^{1}$~Cnc~e.
\label{super}}
\tablehead{
\colhead{parameter} & {value} & {notes}
}
\startdata
Period (days) & $0.736546\pm0.000003$ & $$ \\
K (m\,s$^{-1}$) & $6.30\pm0.21$ & $$ \\
e & $0.0$ & (fixed) \\
$\omega$ (deg) & $90$ & (fixed) \\
T$_{\rm transit} (JD)$ & $55568.011\pm0.008$ & 55568.03469 \\
RMS (m\,s$^{-1}$) & 6.09 & HJST/Tull ($\gamma =-22574.1\pm0.7$\,m\,s$^{-1}$ )\\
RMS (m\,s$^{-1}$) & 5.62 & HET/HRS ($\gamma = 28394.0\pm0.08$\,m\,s$^{-1}$ ) \\
RMS (m\,s$^{-1}$) & 3.25 & Keck/HIRES ($\gamma = 14.4\pm0.5$\,m\,s$^{-1}$ )\\
RMS (m\,s$^{-1}$) & 7.55 & Lick/Hamilton ($\gamma = 14.2\pm0.5$\,m\,s$^{-1}$ )\\
\hline
mass ($M_{\oplus}$) & $8.37\pm0.38$ & $i=82.5$\\
$\rho$ (g\,cm$^{-3}$) & $4.50\pm0.20$ & \\
\hline
\enddata
\end{deluxetable}

\begin{deluxetable}{lrrrr}
\tablecolumns{5}
\tablewidth{0pt}
\tablecaption{Keplerian Orbital Solutions for the 4 giant planets in the $\rho^{1}$~Cnc system.
\label{4planets}}
\tablehead{
\colhead{parameter} & {planet b} & {planet c} & {planet f} & {planet d}
}
\startdata
Period (days) & $14.651\pm0.0001$ & $44.38\pm0.007$ & $261.2\pm0.4$ & $4909\pm30$ \\
K (m\,s$^{-1}$) & $71.11\pm0.24$ & $10.12\pm0.23$ & $6.2\pm0.3$ & $45.2\pm0.4$ \\ 
e & $0.004\pm0.003$ & $0.07\pm0.02$ & $0.32\pm0.05$ & $0.02\pm0.008$ \\ 
$\omega$ (deg) & $110\pm54$ & $356\pm22$ & $139\pm8$ & $254\pm32$ \\
T$_{\rm per}$ (BJD) & $53035.0\pm2.2$ & $53083\pm3$ & $51878\pm5$ & $53490\pm437$ \\
\hline
$M \sin i$ ($M_{\rm Jup}$) & $0.80\pm0.012$ & $0.165\pm0.0054$ & $0.172\pm0.008$ & $3.53\pm0.08$ \\  
a (AU) & $0.1134\pm0.0006$ & $0.237\pm0.0013$ & $0.77\pm0.005$ & $5.47\pm0.06$ \\
\hline
\enddata
\end{deluxetable}

\begin{figure}
\includegraphics[angle=270,scale=.62]{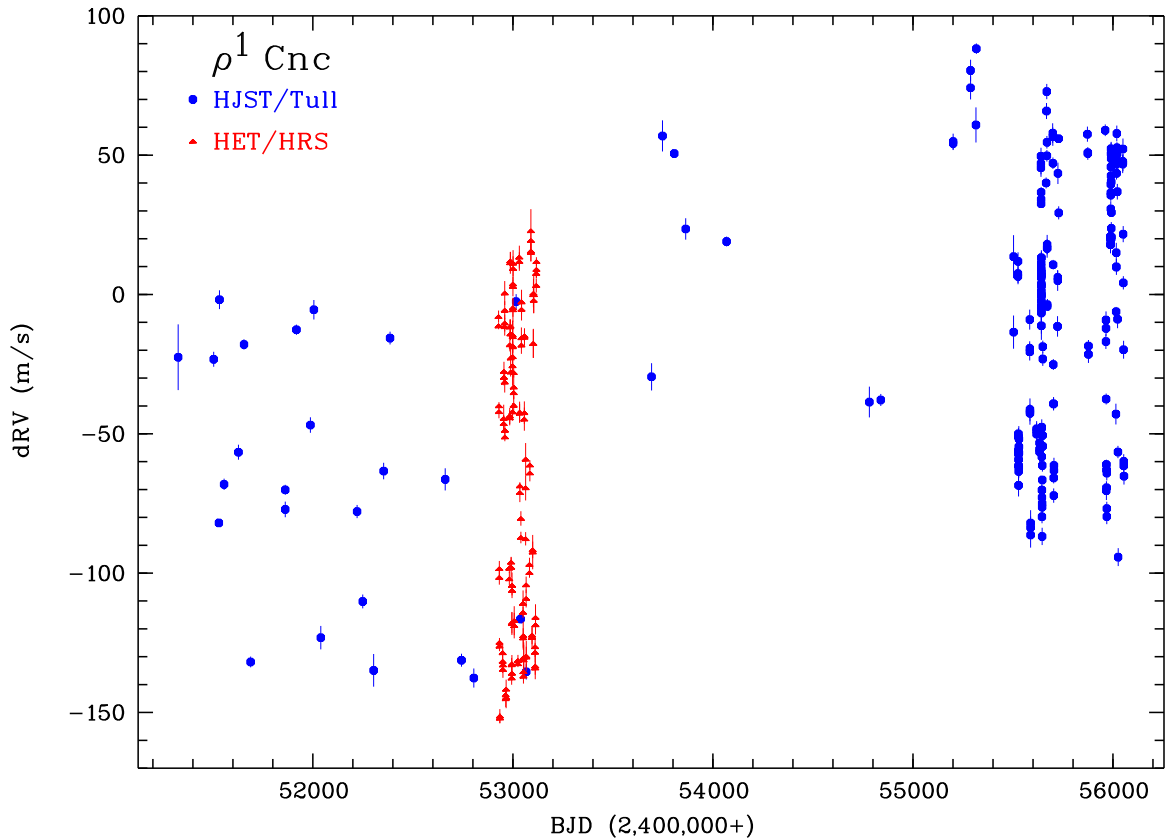}
\caption{13 years of precise RV measurements of $\rho^{1}$~Cnc by the McDonald Observatory planet search. The filled
(blue) circles are the HJST/Tull RV data and the filled (red) triangles represent the HET/HRS results. 
The two dense clusters of points are the high cadence campaigns we performed in 2003/2004 with the HET/HRS and starting
in 2010 with the HJST/Tull. We subtracted from
both data sets the arbitrary, best-fit RV zero-points (HJST: $\gamma=-22574.1$\,m\,s$^{-1}$, HET: $\gamma= 28394$\,m\,s$^{-1}$)
from the joint Keplerian orbit fit (see section\,\ref{gf}). The large scatter of these data are almost entirely due to the
Keplerian motion due to the planets in the $\rho^{1}$~Cnc system.
\label{rvs}}
\end{figure}

\end{document}